\documentclass[prl,showpacs,superscriptaddress,twocolumn]{revtex4}

\usepackage{graphicx}
\usepackage{color}
\usepackage{latexsym}
\usepackage{amsmath}
\usepackage{amsthm}
\usepackage{amsfonts}
\usepackage{amssymb}
\usepackage{bm}

\newcommand{\rem}[1]{}
\newcommand{\refe}[1]{(\ref{#1})}

\newcommand{\refE}[1]{Eq.~(\ref{#1})}
\newcommand{\beq}{\begin{equation}}
\newcommand{\eeq}{\end{equation}}
\newcommand{\beqa}{\begin{eqnarray}}
\newcommand{\eeqa}{\end{eqnarray}}
\newcommand{\eps}{\varepsilon}

\def\Tr{\mathrm{Tr}}
\def\tr{\mathrm{tr}}
\def\sign{\mathrm{sign}}

\def\PauliNambu{\tau}    
\def\PauliSpin{\sigma}   

\begin{document}
\title{
Distribution function of persistent current
}

\author{M. Houzet}

\affiliation{SPSMS, UMR-E 9001, CEA-INAC/UJF-Grenoble 1, F-38054 Grenoble, France}

\date{\today}

\begin{abstract}

We introduce a variant of the replica trick within the nonlinear sigma model that allows
calculating the distribution function of the persistent current. 
In the diffusive regime, a Gaussian distribution is derived. This result holds in the 
presence of local interactions as well. Breakdown of the Gaussian statistics is predicted
for the tails of the distribution function at large deviations.

\end{abstract}

   		
\maketitle

{\it Introduction.~-}
A striking manifestation of quantum mechanics in the mesoscopic 
physics of electrons is that an equilibrium persistent current (PC) can flow
in normal-metallic rings threaded by a magnetic 
flux~\cite{PC}. In the diffusive regime, this property 
arises from a flux-dependent interference
contribution to the electron density of states~\cite{DOS} that survives
in presence of the impurity-induced static potential disorder.
The amplitude of PC is quite small and varies strongly from
sample to sample: The theory for non-interacting electrons~\cite{cheung89,riedel} 
predicts that the ensemble average vanishes, while the typical
amplitude is $\Phi_0$-periodic ($\Phi_0=h/e$ is the flux quantum) and scales as 
$I_\mathrm{typ}\sim e/\tau_D$, 
where $\tau_D$ is the diffusion time along the ring. 
Thus, $I_\mathrm{typ}\sim 1\mathrm{nA}$ in
micrometer-size rings made of conventional metals. 
Electron-electron interactions were predicted to induce a 
$\Phi_0/2$-periodic average current of order $I_\mathrm{av}\sim\lambda_\mathrm{eff} e/\tau_D$,
where $\lambda_\mathrm{eff}$ is an effective coupling constant~\cite{ambegaokar1}. In superconducting rings,
this would yield a diamagnetic average current due to superconducting
fluctuations well above the superconducting critical temperature~\cite{ambegaokar2}.

Early experiments measured the PC 
by detecting the small magnetic field that they produce with
superconducting quantum interference devices (SQUIDs).
Both the $\Phi_0/2$-periodic average current
in large ensembles of rings~\cite{levy,deblock} and $\Phi_0$-periodic current in single rings and
small ensembles of rings~\cite{chandrasekhar,mailly,rabaud,jariwala} were recorded;
a low-flux diamagnetic response was observed in~\cite{jariwala,deblock}.
However, it was not always possible to reconcile their amplitude 
with the theories described above. 
This could be due to subtle effects related to 
canonical vs.grand canonical averaging in isolated 
rings~\cite{bouchiat89},
magnetic impurities~\cite{bary-soroker}, 
or the great sensitivity of the PC to its electromagnetic environment~\cite{kravtsov}.
A recent experiment on several rings addressed separately with a scanning SQUID microscope 
showed however a good agreement with the non-interacting theory
for the typical current and no sign of average current~\cite{bluhm2009}.

An experimental breakthrough was made recently by measuring PCs
with high-precision cantilever torque magnetometry~\cite{yale}. 
Notably, the technique allows better sensitivity and works under large magnetic fields 
(compared with the SQUID technique). The PC could thus be recorded over a huge number 
of flux periods. The amplitude of the typical PC was found to be in good agreement 
with the prediction for non-interacting electrons and no average current was detected. 

The aim of the present work is to address the distribution function of PC
that seems within reach of the new experimental technique.
We demonstrate that in the diffusive regime the statistics is Gaussian.
It justifies characterizing the PC with its first two cumulants only.
The Gaussian distribution carries on in the presence of local interactions as well.
To derive the result, we make use of a replica trick in the nonlinear sigmal model 
that allows obtaining the distribution function at once.
It provides an alternative to the evaluation of all cumulants order 
by order \cite{smith1992,Bussemaker1997}. It
could serve as a useful starting point to address questions such as 
canonical vs.~grand canonical averaging, localization 
and cross-over to non-Gaussian statistics.

{\it Gaussian distribution for the persistent current of non-interacting electrons.~-}
The PC flowing in a metallic ring
pierced by a magnetic flux $\Phi$,
\beq
I(\Phi)=-\frac{\partial  F}{\partial \Phi},
\label{eq:PC}
\eeq
is related to the free energy $F(\Phi)=-kT\ln Z_\Phi$, where
$Z_\Phi$ is the partition function.
The PC distribution function at given $\Phi$ is the probability 
density for $I(\Phi)$ to be equal to ${\cal I}$,
\beq
\label{eq:PI}
P({\cal I})=
\langle\delta \left({\cal I}-I(\Phi)\right)\rangle,
\eeq
where the brackets denote an ensemble average over different disorder 
configurations in the ring.
Using the identity $\delta(s)=\int (dx/2\pi)e^{ixs}$ and
the definition of the derivative, we express \refE{eq:PI} as
\beq
\label{eq:PI2}
P({\cal I})
=
\lim_{\Phi'\rightarrow\Phi}\int \frac{dx}{2\pi}e^{i{\cal I} x}
\langle Z_\Phi^nZ_{\Phi'}^{n'}\rangle ,
\eeq  
where $n=-n'=-ixkT/(\Phi-\Phi')$.

The disorder-averaging for free fermions in \refE{eq:PI2} 
can be performed with a variant of the replica trick~\cite{replicas}
in the fermionic non-linear sigma model~\cite{Wegner,Fin}. 
For this, we consider that the system is formed of $n$ replicas with flux $\Phi$ 
and $n'$ replicas with flux $\Phi'$
($n$ and $n'$ integers) and we evaluate
\beq
\label{eq:PartFunc}
    \langle Z_\Phi^nZ_{\Phi'}^{n'} \rangle
    =
    \int {\cal D} Q   e^{-S[Q]}.
\eeq
Here, $Q$ is a matrix field acting
in the direct product of
the replica space of dimension $n+n'$,
infinite Matsubara energy space,
two-dimensional Gorkov-Nambu space (Pauli matrices $\PauliNambu_\alpha$),
and two-dimensional spin space (Pauli matrices $\PauliSpin_k$).
The $Q$ matrix obeys the nonlinear constraint
$Q^2=1$ and the charge conjugation symmetry
$Q=\overline{Q}
\equiv \PauliNambu_1 \PauliSpin_2 Q^T \PauliSpin_2 \PauliNambu_1$,
where $Q^T$ stands for the full matrix transposition.
The action of the model is~\cite{Fin,Efetov-book}
\beq
S[Q]
=
\frac{\pi\nu}{8}
\int d\bm{r}
\Tr \left[ D({\bm \partial} Q)^2 - 4
	\epsilon \PauliNambu_3 Q 
\right]
.
\label{eq:action}
\eeq
Here, $\nu$ is the single-particle density
of states per spin, 
$D$ is the diffusion coefficient in the metal, 
$\bm{\partial}=\bm{\nabla}+(ie/\hbar)[\bm{A}\PauliNambu_3,.]$
includes the effect of a vector potential associated with 
the magnetic field $\bm{B}=\mathrm{rot}\bm{A}$ (different in each replica),
$\epsilon$ is a fermionic
Matsubara energy, and the trace `$\Tr$' is taken over all spaces
of the $Q$ matrix. 

Let us consider a quasi one-dimensional circular ring with length $L=2\pi R$
($R$ is the radius of the ring). The circular gauge
$\bm{A}=(\Phi/L)\bm{u}$, where $\bm{u}$ is a unitary vector
normal to the ring, is used. We introduce a coordinate $y$ along the ring
measured in units of $L$, a flux vector $\hat{\phi}=\{\phi_a\}_a$
in replica space measured in units of $\Phi_0$, 
with components $\phi_a=\phi\equiv \Phi/\Phi_0$ 
($\phi_a=\phi'\equiv \Phi'/\Phi_0$) for $1\leq a\leq n$ ($n<a\leq n+n'$),
and $\varepsilon=\epsilon/E_c$, 
where $E_c=\hbar D/R^2$ is the Thouless energy related to diffusion time
($E_c=4\pi^2\hbar/\tau_D$). 
Then, \refE{eq:action} reads
\beq
\label{eq:action2}
S[Q]
=
\frac{E_c}{32\pi\delta}
\int_0^1 d y
\Tr \left[ (\partial_y Q)^2 - 16\pi^2
	\eps \PauliNambu_3
	 Q 
\right]
,
\eeq
where $\delta$ is the mean level spacing in the ring
and $\partial_y=\nabla_y+2\pi i[\hat{\phi} \PauliNambu_3,.]$.
The single-valuedness of the $Q$ field fixes the boundary conditions: 
$Q(y=0)=Q(y=1)$ and $\nabla_y Q(y=0)=\nabla_y Q(y=1)$. 

In the metallic regime, the ring's conductance
measured in units of the conductance quantum, $g=E_c/(2\pi\delta)$, is large.
Thus, we can evaluate \refE{eq:PartFunc} within the saddle-point approximation.
The $Q_0$ field that minimizes the action (\ref{eq:action2}) 
is proportional to unity in spin 
and replica spaces, and diagonal in Matsubara space with value 
$Q_{0}(\eps)=\PauliNambu_3\sign(\eps)$.
The action at the saddle point is $(n+n')S_0$ where
$S_0=-(2\pi E_c/\delta)\sum_\eps|\eps|$; it does not depend on the flux.

\begin{widetext}
In order to study fluctuations near this saddle point we write
matrices close to $Q_0$ as
\beq
\label{Q-W}
    Q = Q_0 (1+W+W^2/2+\dots) .
\eeq
The constraints on the $Q$ field imply that $\{Q_0,W\}=0$ and
$W=-\overline{W}$; the requirement of convergency of the sigma model on the 
perturbative level implies that $W^\dagger=-W$.
Then, we decompose the $W$ field into its elements in Nambu and spin spaces,
and its Fourier components: 
\beq
    W_{\eps a ,\eps' b}(y)
	=
	\sum_{p=-\infty}^\infty
	\sum_{k=0}^3
	\left(
		\begin{array}{cc}
			W_{\eps a ,\eps' b}^{k(p)} & \tilde W_{\eps a ,\eps' b}^{k(p)} \\
			-\tilde W_{\eps' b ,\eps a}^{k(-p)*} & \varsigma_k W_{-\eps' b ,-\eps a}^{k(p)}
		\end{array}
	\right)_\mathrm{Nambu}
	\otimes \PauliSpin_k
	e^{2\pi i p y }
	,
\label{eq:W}
\eeq
where $\varsigma_k=\pm$ for $k=1,2,3$ and $k=0$, respectively, and the components 
$W_{\eps a ,\eps' b}^{k(p)}$ ($\tilde W_{\eps a ,\eps' b}^{k(p)}$) 
only exist at $\eps\eps'<0$ ($\eps\eps'>0$).
An independent set of complex integration variables is then 
obtained with 
$W_{\eps a ,\eps' b}^{k(p)}$ and $\tilde W_{\eps a ,\eps' b}^{k(p)}$
at $a>b$, and  
$W_{\eps a ,\eps' a}^{k(p)}$ and $\tilde W_{\eps a ,\eps' a}^{k(p)}$
at $\eps>0$.

Expanding the action (\ref{eq:action2}) near the saddle point up to quadratic terms
in $W$, one finds:
\beq
\label{S2-QD}
    S^{(2)}[W]
    =
    \frac{\pi E_c}{2\delta}
    \sum_{\eps\eps'}
    \sum_{ab}
    \sum_k
	\sum_p
\left\{
\left[|\eps|+|\eps'|+(p+\phi_a-\phi_b)^2\right]
|W_{\eps a,\eps'b}^{k(p)}|^2
+\left[|\eps|+|\eps'|+(p+\phi_a+\phi_b)^2\right]
|\tilde W_{\eps a,\eps'b}^{k(p)}|^2\right\}
\eeq
\end{widetext}
The Gaussian integration over the $W$ field is then straightforward and yields:
\beq
\label{eq:Z}
    \langle Z_\Phi^nZ_{\Phi'}^{n'} \rangle
    =
    e^{-n^2\Xi (\phi,\phi)-n'^2\Xi (\phi',\phi')-2nn'\Xi (\phi,\phi')},
\eeq
where we have omitted a factor
which is equal to 1 in the replica limit $n'=-n$, and
\beq
\Xi (\phi,\phi')=4\sum_p\sum_{\eps,\eps'>0}\sum_{s=\pm}
\ln\left[\eps+\eps'+(p+\phi-s\phi')^2\right].
\eeq

The replica trick now consists in assuming
that Eq.~(\ref{eq:Z}) can be analytically continued to pure
imaginary variables $n$ and $n'$, with $n'=-n$. Then, 
inserting Eq. (\ref{eq:Z}) into (\ref{eq:PI2}), we find
\beq
\label{eq:distribution}
P({\cal I})
=
\int \frac{dx}{2\pi}e^{i{\cal I} x}
e^{-x^2 I_\mathrm{typ}^2/2}
=
e^{-{\cal I}^2/(2 I_\mathrm{typ}^2)}
/{\sqrt{2\pi I_\mathrm{typ}}},
\eeq  
where $I_\mathrm{typ}^2=-2(kT/\Phi_0)^2\partial^2\Xi (\phi,\phi')/(\partial \phi\partial \phi')|_{\phi'=\phi}$.
That is, the distribution function of the PC
is Gaussian with a zero mean value and typical value 
$I_\mathrm{typ}$. Using the Poisson
summation rule, we can convert the sum over $p$ in $\Xi$ into an integral
and finally obtain, at zero temperature,
\beq
\label{eq:variance}
I_\mathrm{typ}^2(\Phi)
=
\frac{24 E_c^2}{\pi^2\Phi_0^2}
\sum_{q=1}^\infty
\frac{1}{q^3}
\sin^2\left(2\pi q \Phi/\Phi_0\right).
\eeq
Thus, the PC 
has typical amplitude $E_c/\Phi_0\sim e/\tau_D$. 
The distribution function (\ref{eq:distribution}) is in agreement 
with the known results~\cite{riedel} for the average
current $\langle I(\Phi) \rangle$ and its cumulant
$\langle\langle I(\Phi) I(\Phi') \rangle\rangle
\equiv
\langle I(\Phi) I(\Phi') \rangle-
\langle I(\Phi) \rangle
\langle I(\Phi') \rangle$ in 
non-interacting diffusive rings,
with $I_\mathrm{typ}=\langle\langle I(\Phi)^2 \rangle\rangle^{1/2}$~\cite{note}.

This section contains the main result of this article, Eq.~\refe{eq:distribution}. In the following
we illustrate several directions where it can be extended.

{\it Distribution function for the harmonics.~-}
It may be more convenient experimentally to characterize the
flux-current relation by its harmonic content. In this section, 
we show that the distribution function for the harmonics in the diffusive
regime is also Gaussian.

Due to time-reversal symmetry and flux-periodicity, 
the current-flux relation,
\beq
\label{eq:curr-phase}
I(\Phi)
=
\sum_{q=1}^\infty
I_q \sin 2\pi q\phi,
\eeq
is fully characterized by its harmonics $I_q$.
The distribution function for the harmonics,
\beq
\label{eq:distrib-harm}
P_q({\cal J})=\langle\delta ({\cal J}-I_q)\rangle,
\eeq
can also be determined with a replica trick.
Indeed, using Eqs.~\refe{eq:PC}, \refe{eq:curr-phase}, and integrating by parts, we first note that
$I_q=(8\pi q/\Phi_0) \int_0^{1/2}d\phi \cos(2\pi q \phi) F(\phi)$.
By definition of the integration, it also reads:
\beq
\label{eq:harmonics}
I_q=\lim_{N\rightarrow\infty}(4\pi q /N\Phi_0)\sum_{\ell=1}^N \cos(2\pi q \phi_\ell) F(\phi_\ell),
\eeq
where $\phi_\ell=\ell/(2N)$.
Now, inserting the representation of the delta-function and 
Eq.~\refe{eq:harmonics} into \refe{eq:distrib-harm}, 
we find
\beq
P_q({\cal J})
=
\lim_{N\rightarrow\infty}
\int\frac{dx}{2\pi}
e^{i{\cal J} x}
\langle
\prod_{\ell=1}^N
Z_{\phi_\ell}^{n_\ell}
\rangle,
\eeq
where $n_\ell=i 4\pi x q kT\cos(2\pi q \phi_\ell)/(\Phi_0 N)$.

The average over the disorder can also be performed within the fermionic sigma model
by considering that the system is formed of $n_\ell$ replicas ($n_\ell$ integer) with flux $\phi_\ell$
($1 \leq \ell\leq N$). In the saddle point approximation, one would find
as a generalization of Eq.~(\ref{eq:Z}):
\beq
\langle
\prod_\ell
Z_{\phi_\ell}^{n_\ell}
\rangle
=\exp\left[-\sum_{\ell\ell'}n_\ell n_{\ell'}\Xi(\phi_\ell,\phi_{\ell'})\right].
\eeq
Taking the replica limit, one again obtains a Gaussian distribution 
$P_q({\cal J}) \sim \exp(-{\cal J}^2/2\langle\langle I_q^2 \rangle\rangle)$
for the harmonics, with zero average value and variance 
\beqa
\langle\langle I_q^2 \rangle\rangle
&=&-
\frac{32k^2T^2}{\Phi_0^2} \int_0^{1/2}
d\phi d\phi'\sin(2\pi q \phi)\sin(2\pi q \phi')
\nonumber \\
&&\times
\frac{\partial^2\Xi (\phi,\phi')}{\partial \phi\partial \phi'}
.
\eeqa 
In particular, $\langle\langle I_q^2 \rangle\rangle=96E_c^2/(\pi^2\Phi_0^2q^3)$
at $T=0$, in agreement with Eq.~\refe{eq:variance}.

{\it Interactions.~-}
The effect of electron-electron interactions can also be taken into account.
To be specific, we consider the case of attractive, local pairing between electrons with
opposite spins that was theoretically debated after the early experiments on PC. 
Then, the action \refe{eq:action} should be supplemented with 
an interaction term~\cite{Fin},
\beqa
\label{eq:action_int}
S_\mathrm{int}[Q]
&=&
-\frac{\nu\lambda\pi^2k^2T^2}{16}
\sum_a
\int d\bm{r} d\tau
\left[(\tr\PauliNambu_1 Q_{\tau a ,\tau a})^2
\right.
\nonumber \\
&&\left.+(\tr\PauliNambu_2 Q_{\tau a ,\tau a})^2]
\right],
\eeqa
where $\lambda$ is the Bardeen-Cooper-Schrieffer coupling constant, $\tau$ is imaginary time,
and the trace `$\tr$' is taken over spin and Nambu spaces.

Above the superconducting critical temperature, the action
can be evaluated in the Gaussian approximation 
near the metallic saddle point $Q_0$.
For the one-dimensional ring, Eq.~\refe{eq:action_int}
results in an interacting contribution adding to \refe{S2-QD}:
\beq
\label{eq:ZZInt}
S_\mathrm{int}^{(2)}[W]
=-\frac{4\lambda\pi^2T}{\delta}
\sum_{p,\omega, a}\sum_{\eps,\eps'>0}
\tilde W^{0(p)}_{\eps a, \eps+\omega a}
\tilde W^{0(p)*}_{\eps' a, \eps'+\omega a}.
\eeq
where $\omega$ is a bosonic Matsubara energy (also measured in units of $E_c$).
Gaussian integration over the $W$ field including Eqs.~\refe{S2-QD}, \refe{eq:ZZInt} can be 
performed; it yields
\beq
\label{eq:ZZInt2}
	\langle Z_\Phi^nZ_{\Phi'}^{n'} \rangle_\mathrm{\lambda}
    =
   \langle Z_\Phi^nZ_{\Phi'}^{n'} \rangle_{\lambda=0} 
    e^{-n\Xi_\mathrm{int}(\phi)-n'\Xi_\mathrm{int}(\phi')},
\eeq
where
\beq
\Xi_\mathrm{int}(\phi)=\!
\sum_{p\omega}
\ln\left[
1
-\!\!\!\!\!\!
\sum_{|\omega|<\eps<\frac{\Omega}{E_c}}
\!\!\!\frac{(4\lambda\pi kT/E_c)}{2\eps-|\omega|+(p+2\phi)^2}
\right].
\eeq
Here,  $\Omega$ fixes the energy bandwidth around the Fermi level over which 
pairing is effective. By introducing the critical temperature
$T_c\simeq (1.14\Omega/k) e^{-1/\lambda}$, one gets
\beqa
\Xi_\mathrm{int}(\phi)=
\sum_{p\omega}
\ln\left[
\ln\frac{T}{T_c}
+\psi\left(\frac{1}{2}
+\frac{|\omega|+(p+2\phi)^2}{4\pi k T/E_c}
\right)
\right.
\nonumber\\
\left.
-\psi\left(\frac{1}{2}\right)
\right].
\eeqa
where $\psi$ is the digamma function. Inserting Eq.~\refe{eq:ZZInt2}
into \refe{eq:PI2}, one again finds that the distribution function 
for PC is Gaussian, with average value
$\langle I(\Phi) \rangle= -(kT/\Phi_0)\partial \Xi_\mathrm{int}/\partial\phi$ 
and the same variance as in the non-interacting case.
The average current was discussed in Ref.~\cite{ambegaokar2}, 
it is $\Phi_0/2$-periodic with amplitude 
$I_\mathrm{av}\sim \lambda_\mathrm{eff} e/\tau_D$
where $\lambda_\mathrm{eff}\sim \ln^{-1}(E_c/kT_c)$
at $T_c < T \ll E_c$.

{\it Discussion.~-}
We first note that spin and orbital effects, such as 
the penetration of the magnetic field 
within rings with finite thickness~\cite{ginossar} 
are important for a quantitative comparison 
with the experiment~\cite{yale}.
Taking these effects into account within our formalism can be done easily; 
it would not change the prediction of a Gaussian distribution 
in the diffusive regime.

On the other hand, the Gaussian statistics clearly fails 
in the insulating regime, at $g < 1$.
Actually, its alteration is expected already at large, but finite $g$, 
in relation with the Anderson localization phenomenon.
A similar question on the statistics of the -- dissipative -- conductance 
of diffusive wires was addressed~\cite{AKL}. 
Log-normal tails in the probability distribution
were predicted at large deviations from the average conductance. 
However, the present case differs by the fact that PC is a 
thermodynamic quantity.
 
To estimate the range of validity of the Gaussian statistics for PC,
we expand the action \refe{eq:action} in vicinity of 
the metallic saddle point $Q_0$ up to fourth order terms in the 
field $W$. Then, we evaluate the generated terms perturbatively
with the Gaussian action. As a result, we found that the leading correction 
to the integrand in Eq.~\refe{eq:distribution} arises in order 
$\propto x^3 I_\mathrm{typ}^3/g$, consistent with the recently
derived third order cumulant~\cite{danon}.
The same way, we also obtain  that $n$-th order cumulants
scale as  $I_\mathrm{typ}^n/g^{n-2}$ at $n\geq 3$.
Subsequently, this implies that the Gaussian distribution is not reliable
at large deviations, when $|{\cal I}|\gtrsim g^{1/3} I_\mathrm{typ}\gg I_\mathrm{typ}$. 
A more detailed investigation of the behaviour of $P({\cal I})$ at large deviations 
is left for future study.

In the absence of interactions, the average current vanishes. 
However, this result is an artefact of the grand canonical
averaging tacitly performed here. When (canonical) averaging is done with keeping the
number of electrons constant in the ring, a small, but finite, average current 
$I_\mathrm{av}\sim I_\mathrm{typ}/g$ is obtained~\cite{Iav}.
Including this effect in the framework of this article
remains an open question.

{\it Conclusion.-}
The persistent current has been mostly characterized by its first two cumulants. 
Here, we proposed a replica trick allowing to calculate at once all the cumulants 
or, equivalently, the complete distribution function. We mostly applied
this trick to the diffusive regime, when the statistics is Gaussian 
and higher order cumulants are negligible.
We believe that the trick could be extended to regimes
where the Gaussian statistics breaks down. 

The replica trick introduced in this paper can be applied to 
the evaluation of the probability distribution of
other thermodynamic quantities. For instance, 
the nonlinear sigma-model was used to calculate the mesoscopic 
fluctuations of the supercurrent in metallic Josephson junctions~\cite{sns}. 
We would easily find that the statistics of 
the supercurrent is also Gaussian in the diffusive regime.

\acknowledgements
I am grateful to L.~Glazman for drawing my interest to the question 
addressed in this article, to him, J.~Meyer, and G. Montambaux for 
many useful discussions, and
to the Nanosciences Foundation of Grenoble for support.


\begin{thebibliography}{99}
\bibitem{note}
	Alternatively to the derivation in the main text, 
	one can insert the standard replica trick 
	$F=-kT\lim_{n\rightarrow 0}(Z^n-1)/n$ into 
	Eq.~(\ref{eq:PC}) and relate the two first cumulants
	to Eq.~(\ref{eq:Z}). Then it is found that the average current vanishes 
	and
$
\langle\langle I(\Phi) I(\Phi') \rangle\rangle
=
\sum_{q=1}^\infty
(24 E_c^2/\pi^2\Phi_0^2q^3)
\sin(2\pi q \phi)
\sin(2\pi q \phi')
$
at $T=0$.
\end{thebibliography}
\end{document}